\documentclass[prc,aps,onecolumn,showpacs,preprint]{revtex4}
\usepackage{graphicx}

\begin{document}

\title{ Resonance signals in pion-nucleon scattering from speed plots and time-delay}

\author{
R.L. Workman and R.A. Arndt}
\affiliation{
Center for Nuclear Studies,
Department of Physics\\
The George Washington University,
Washington, D.C. 20052}

\date{\today}

\begin{abstract}

We compare a number of methods used to locate
resonances. These include the speed plot, the time-delay
method of Eisenbud and Wigner, the time-delay matrix of
Smith, and a modification proposed by Ohmura. Numerical
results show a consistency not previously reported. One
time-delay method gives the most consistent results, 
supporting the conclusions of a recent theoretical study.

\end{abstract}


\maketitle

Given a set of elastic scattering amplitudes or, for a multi-channel
process, the full S-matrix, the most direct way to locate resonances
involves a search for poles in the complex energy plane. To do this,
one must have an analytic representation that can be continued, correctly
accounting for the branch cuts. If only numerical values of the amplitudes
are available, less sophisticated methods may still applied, at least for
those resonances with a clear resonance signature. 

Here we will compare two such methods $-$ the speed-plot, advocated
by H\"ohler~\cite{hoehler}, and the time-delay, which was first studied for
elastic and multi-channel scattering by Eisenbud and Wigner~\cite{eisenbud, wigner}. 
These methods have recently been examined in terms of their theoretical basis~\cite{tshlee, kelkar2}.
Here we will briefly review the proposed methods and make comparisons based on their
ability to correlate speed or time-delay peaks with resonance energies. 
In doing so, we will gain a better understanding of some puzzling results associated with
previous applications of these methods. One particular time-delay method is advocated in
the recent study of Ref.~\cite{kelkar2}. A comparison of our numerical results appears
to support their conclusions.

In 1948, Eisenbud~\cite{eisenbud} proposed a method for locating resonances,
in both single- and multi-channel scattering, based on the
time-delay associated with a wavepacket. Wigner quoted~\cite{wigner} a time-delay
result in a later paper focused on the restrictions of causality.
The result of Eisenbud was
\begin{equation}
{\Delta t}^{\rm E} \; = \; \hbar {d \over {dE}} [ arg( S-1 ) ],
\end{equation}
which for a single channel, gives the result
\begin{equation}
{\Delta t}^{\rm E} \; = \; \hbar {{d \delta} \over {dE}},
\end{equation}
$\delta$ being the phase shift. The result quoted by Wigner was
larger by a factor of 2. A reason for this factor was
later noted by Wigner in Ref.~\cite{wheeler, helmut}. 

Smith~\cite{smith} derived a time-delay matrix, based on the
flux passing through an interaction region of radius $R$. His result
for the average lifetime of a metastable state due to a collision 
beginning in the $i^{\rm th}$ channel was
\begin{equation}
Q \; = \; -i\hbar {{d S}\over {dE}} S^{\dagger},
\end{equation}
$S$ being the S-matrix including all open channels. Smith then
claimed that his result could be connected to the Eisenbud result,
using the following representation
\begin{equation}
{\Delta t_{ij}}^{\rm S} \; = 
\; {\rm Re} \left[ -i\hbar {( S_{ij}) }^{-1} {{d S_{ij} }\over {dE}} \right] ,
\end{equation}
implicitly attributed to Eisenbud. From the above two 
results, Smith noted that
\begin{equation}
Q \; = \; \sum_j S^*_{ij} S_{ij} {\Delta t_{ij}}^{\rm S} ,
\end{equation}
followed trivially. Notice, however, that Eqs.~(1) and (4) are {\it not} equivalent.

Objections to Smith's time-delay formalism were expressed in
the extensive study of Ohmura~\cite{ohmura} as noted in Ref.~\cite{wheeler}. Ohmura's result
agreed with that of Eisenbud, though this was not apparent as the Eisenbud thesis remained unpublished
and the Eisenbud result was misquoted by Smith. Ohmura further claimed that 
Smith's results could be made more consistent by subtracting 
off a term due to the outgoing unscattered
wave, giving for the average time delay~\cite{ohmura} 
of a wavepacket beginning in the $i^{\rm th}$ channel 
\begin{equation}
< \Delta t_i >_{Av} \; = \; {N_{ii}}^{-1} \;
{\rm Re} \; \left[ -i\hbar \sum_n ( {S_{in}}^* - \delta_{in} ) {{d S_{jn}}\over {dE}} \right] ,
\end{equation}
with
\begin{equation}
N_{ii} \; = \; \sum_j ( {S_{ij}}^* - \delta_{ij} ) ( S_{ij} - \delta_{ij} ).
\end{equation}

Ohmura's result can be found if one starts with the expression
\begin{equation}
2i T^*T \; {d \over {dE}} \{ {\rm arg} T\} =   T^* {{dT}\over {dE}} - T {{dT^*}\over {dE}} ,
\end{equation}
uses Eq.~1, and multiplies by appropriate factors of $\hbar$ and $2i$, to obtain
\begin{equation}
{\Delta t_{ij}}^E S_{ij} {S_{ij}}^* \; = \; 
{\rm Re} \left[ -i\hbar {S_{ij}}^* {{d S_{ij} }\over {dE}} \right] ,
\end{equation} 
for $i \ne j$. This can be compared to Smith's result in Eq.~5. The sum differs in the
$i = j$ term, which is constructed to be consistent with Eq.~1, in the single-channel 
scattering limit, rather than Eq.~4. However, this sum is not a
proper average, since the weights do not add to unity. To have the sum represent the
average of Eisenbud time-delays, the weights must be renormalized, and the factor
$N_{ii}$ in Eq.~7 is added for this purpose. 

In the following numerical comparison, we will be using the partial-wave decomposed
S-matrix. However, the results of Eisenbud were derived assuming
a single dominant partial wave, whereas the Ohmura result was based on 
the use of the full S-matrix~\cite{kelkar2, ohmura}. It is thus surprising how well these
methods work, applied to a set of partial wave amplitudes derived from a fit to data.

In Figs.~1 and 2, we compare these time-delay methods, and the more common speed-plot method, 
applied to amplitudes obtained in a fit to elastic pion-nucleon scattering.    
\begin{figure}[th]
\includegraphics[height=0.6\textwidth, angle=90]{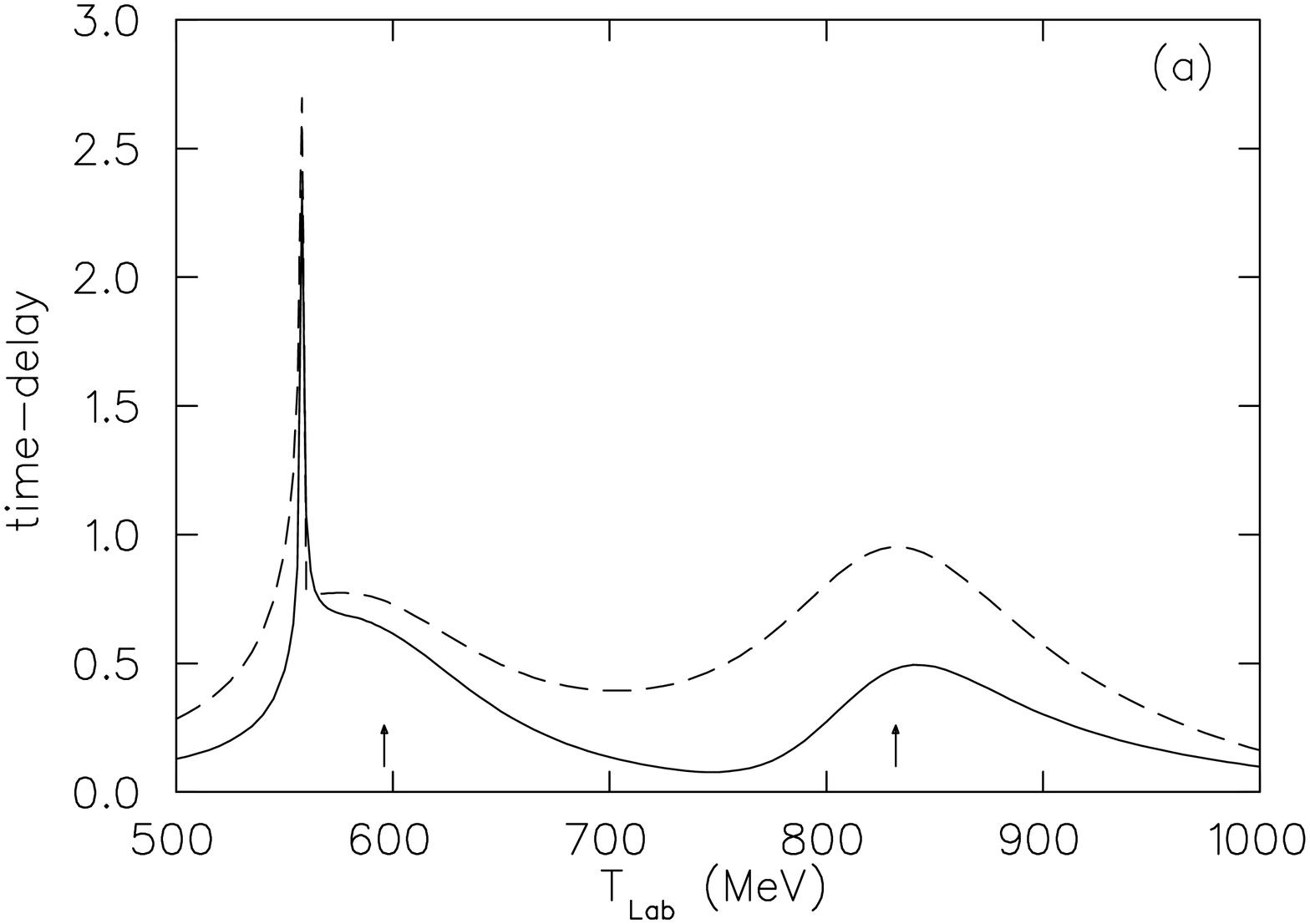}\\
\includegraphics[height=0.6\textwidth, angle=90]{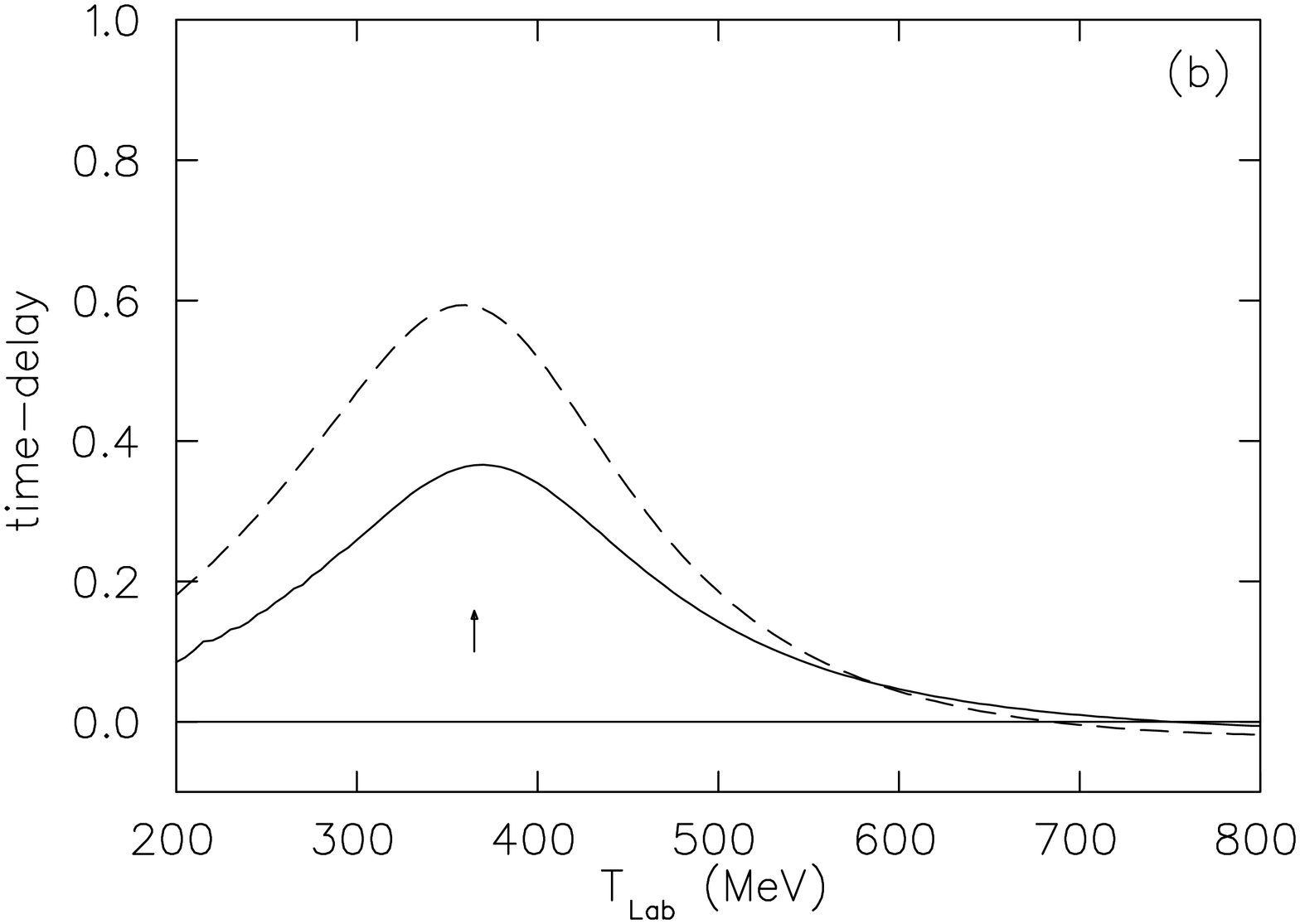}\\
\includegraphics[height=0.6\textwidth, angle=90]{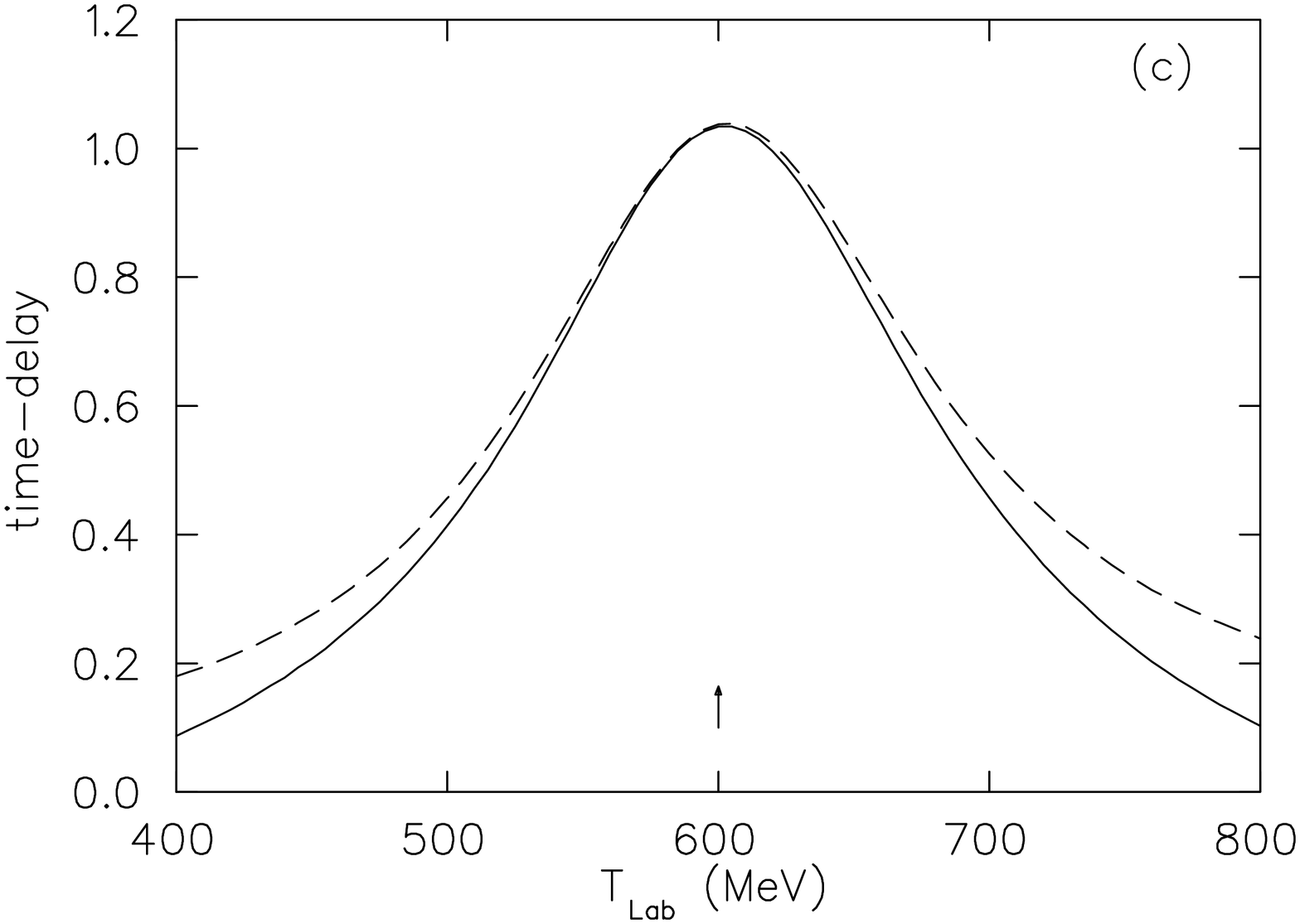}
\caption{Time-delay plots based on the S-matrix approaches of
Smith~\cite{smith} (dashed) and Ohmura~\cite{ohmura} (solid) in
arbitrary units. Arrows denote the real part of resonance pole 
positions for (a) the S$_{11}$, (b) P$_{11}$, and (c) D$_{13}$
partial waves (see text). }
\end{figure}
Figure 1 compares the time delay results of Smith and Ohmura, based on
the scattering S-matrix. In the fit to pion nucleon elastic scattering
and $\eta N$ production data of Ref.~\cite{dac}, a multi-channel Chew-Mandelstam
K-matrix formalism was used. This produced a multi-channel S-matrix, with each 
channel having the required poles and cuts (for $\eta N$, $\pi \Delta$ and $\rho N$).
As the $\pi \Delta$ and $\rho N$ channels were constrained only by the $\pi N$ 
channel inelasticity, the displayed results are a test of the method and would be
improved using a more detailed multi-channel analysis.

Figure 1(a) shows the S$_{11}$ channel, containing the N(1535) and N(1650) resonances,
plus a cusp at the $\eta N$ threshold. Peaking in the time delay is evident at both
resonances, denoted by arrows at the real 
parts of the pole positions~\cite{pole}. Separation of
the N(1535) and the threshold cusp is clear using both forms. In Figure
1(b), results for the P$_{11}$ (Roper) channel are given. Finally,
in Figure 1(c), the D$_{13}$ resonance is shown. In these plots, the normalization
factor $N_{ii}$ is proportional to the imaginary part of the $\pi N$ elastic T-matrix.
For the Roper resonance in particular, dropping this factor causes a significant shift
in the peak position, as the imaginary part of $T_{\pi N}$ is rapidly increasing above
and below the energy corresponding to the real part of the pole position.

\begin{figure}[th]
\includegraphics[height=0.6\textwidth, angle=90]{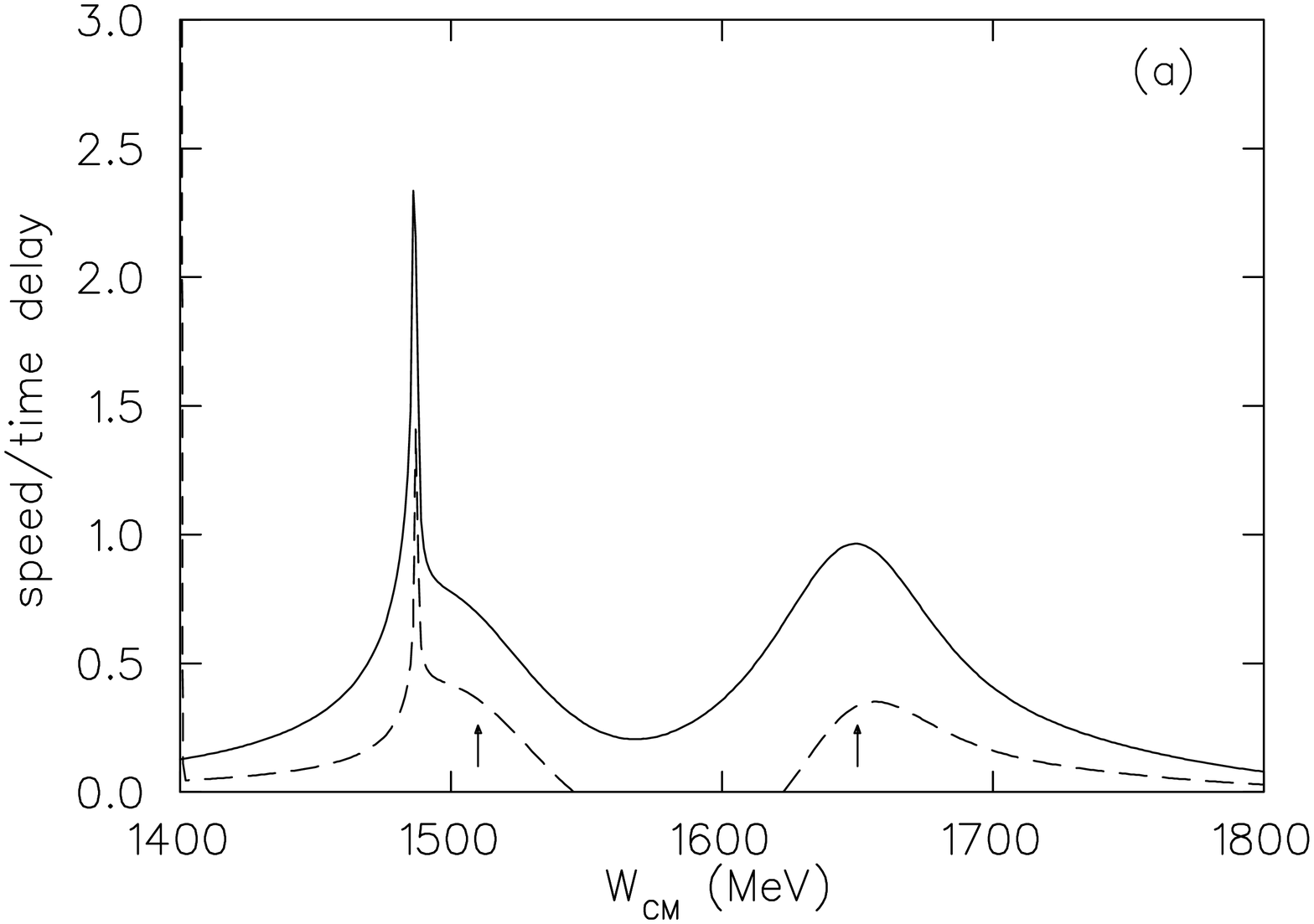}\\
\includegraphics[height=0.6\textwidth, angle=90]{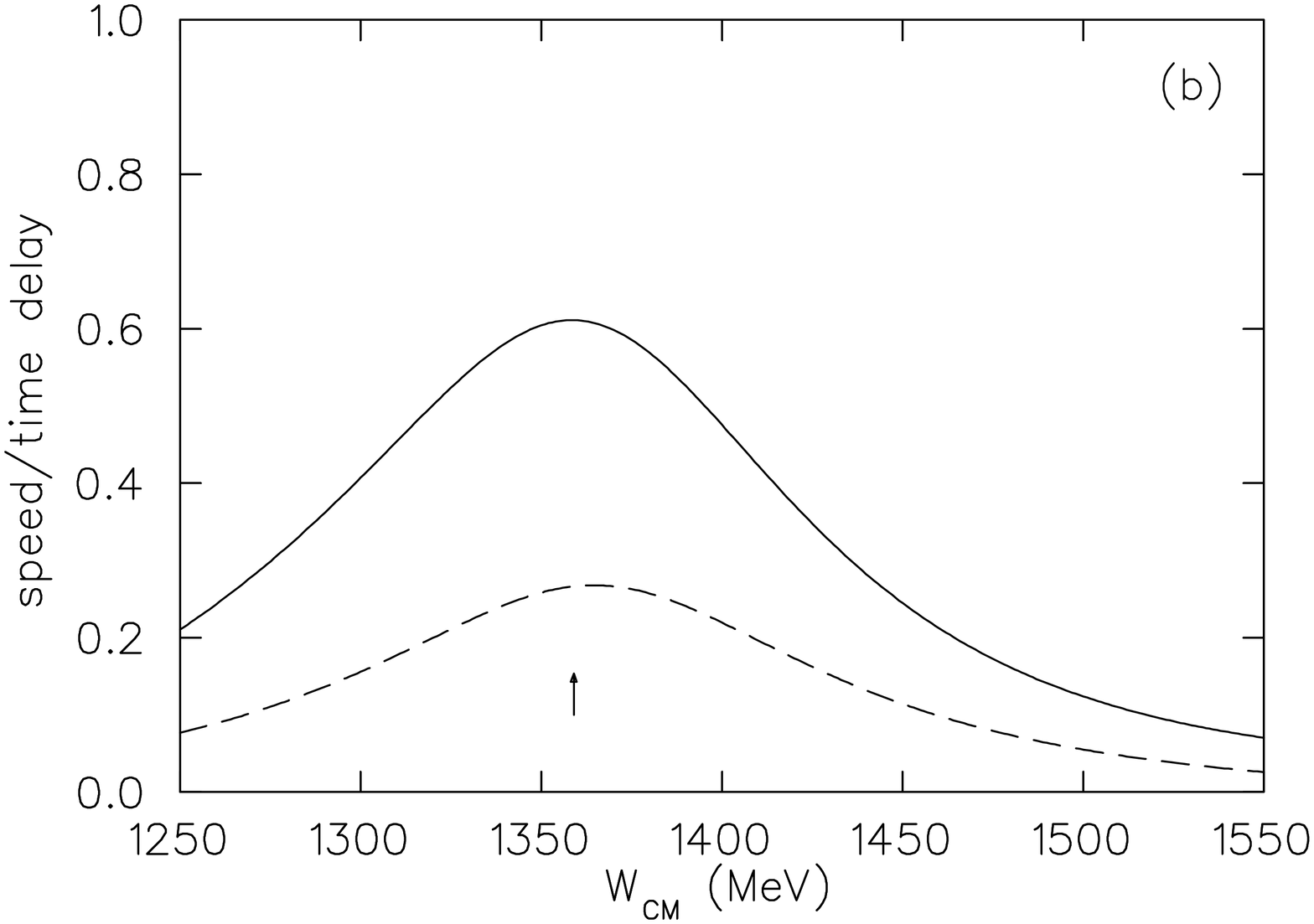}\\
\includegraphics[height=0.6\textwidth, angle=90]{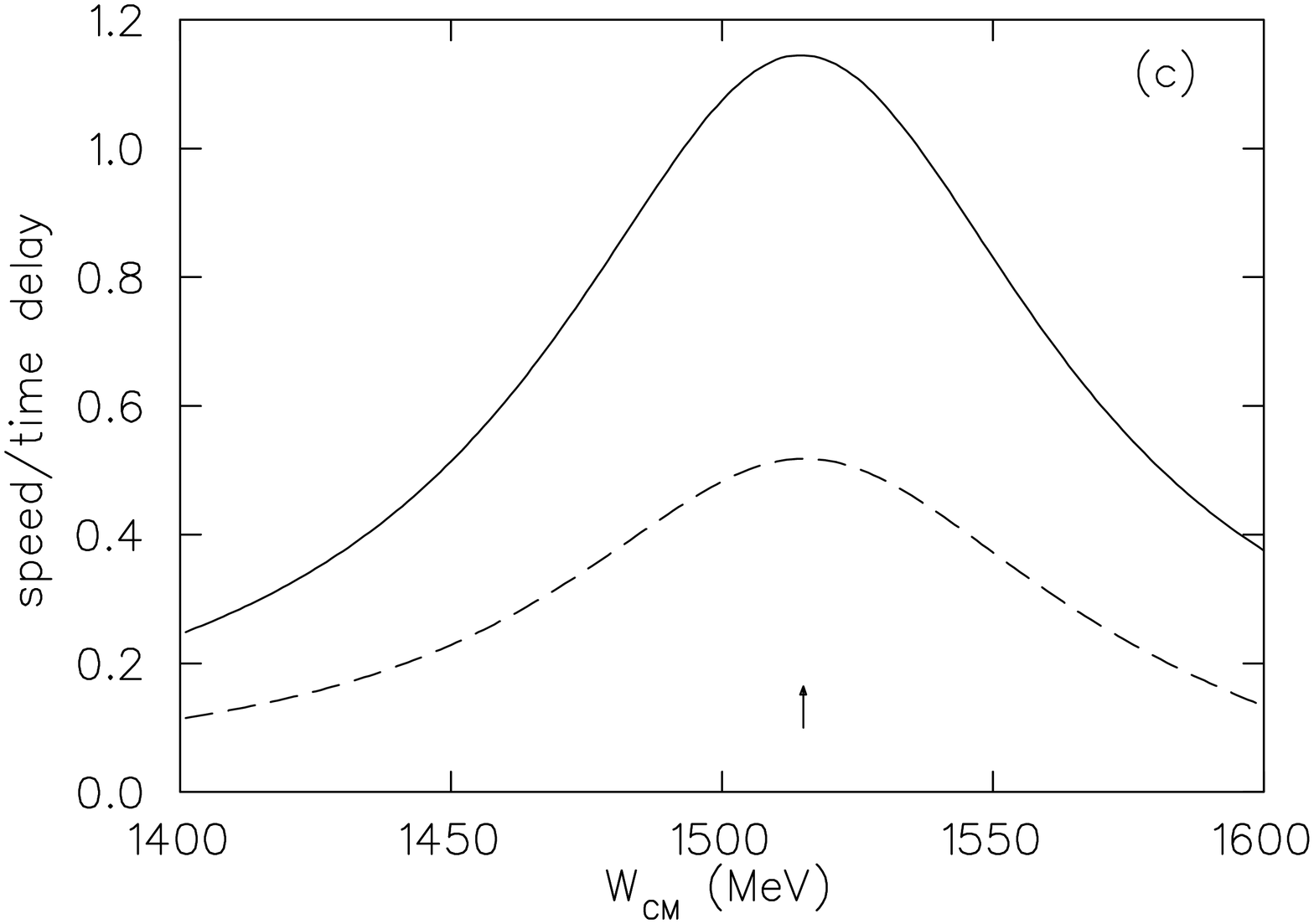}
\caption{ Speed plots (solid) and Eisenbud single-channel 
time-delays (dashed) in arbitrary units. Arrows denote real parts of
resonance pole positions for (a) the S$_{11}$, (b) P$_{11}$, and
(c) D$_{13}$ partial waves (see text).
\label{fig:g2}}
\end{figure}

In Figure 2, we display the speed plot, which is given by the absolute value of
$dT/dW$, and compare this to the single-channel result of Eisenbud, which is
given by the energy derivative of the phase of the T-matrix. Again, for orientation,
we locate the real parts of the associated pole positions. Peaks in the speed plots
again correspond to the pole positions and the Eisenbud time-delay peaks. Separation
of the $\eta N$ cusp and N(1535) resonance is still evident, though not as clear as
in Figure 1. (The region of negative time-delay between the resonances is not plotted.) 
In the study of H\"ohler,  the $\eta N$ threshold and N(1535) resonance
signal were combined into a single peak at the $\eta N$ threshold. 
The separation is visible in our case only if a very fine grid of energies is used.

All methods agree when applied to the elastic P$_{33}$ partial wave and the
$\Delta (1232)$ resonance. All methods find a peak corresponding to the pole position.
We have also applied these methods to the higher partial waves and generally find peaks for all 
PDG 4-star resonances using either the S-matrix methods of Smith and Ohmura or the single-channel
results obtained using the Eisenbud time-delay or speed plot. 
These results are quite different from those of Ref.~\cite{kelkar},
finding no prominent peaks for most of the isospin 3/2 resonances. In that work
Eq.~4 was used for both elastic and inelastic resonances. 

Problems with resonance location and the use of Eq.~4 
are clearly illustrated if we examine the 4-star $\Delta (1950)$ resonance, which produces
a nearly canonical resonance loop in the Argand plane. This resonance has 
$\Gamma_{\pi N} / \Gamma_T$ $<$ 1/2 and therefore the loop passes below the center of the
Argand circle. The four methods applied in Figs.~1
and 2 are compared to Eq.~4 in Fig.~3. All produce peaks at the same 
point, the real part of the energy associated with the pole, except Eq.~4, which produces
a sharp dip closer to resonance mass found in a Breit-Wigner fit to the amplitude.
In Figs.~4 and 5 of Ref.~\cite{kelkar}, resonance positions
were associated with the positive
time-delay shoulder, which is also evident in Fig.~3, prior to the time-advance spike. 

One 4-star resonance, the $\Delta (1620)$ occurring in the S$_{31}$ partial wave, 
breaks the pattern described above. The amplitude, time-delay, and speed plots for this
partial wave are displayed in Fig.~4. In an Argand diagram, this amplitude first moves in a
clockwise direction before beginning the (anti-clockwise) resonance loop. Both matrix methods
and the speed plot produce peaks near the resonance energy. However, the Ohmura result is shifted 
from the pole position. Both of the single-channel time-delay 
results fail to peak at the resonance.  

In summary, we have examined a number of candidate methods for the location of resonances
using either the T-matrix for a single channel, or
the S-matrix accounting for all open channels. These methods generally produce a peak at an energy 
associated with the
real part of the pole position and show remarkable agreement in most cases. 
One method which has been used extensively, based on the energy derivative of the
channel phase shift, does not correlate with the pole position, producing a peak at the 
pole mass of the $\Delta (1232)$, but time-advance dips near the Breit-Wigner masses
of some other more inelastic resonances. The $\Delta (1620)$ resonance appears to provide
the most demanding test (from a well established state). For this state, the matrix methods
and speed plot peak near the resonance energy, the single-channel time-delay methods do not.
Apart from the speed plot, only the S-matrix method of Smith appears to consistently correlate 
with resonance pole positions. This agrees with a recent claim from Kelkar
and Nowakowski~\cite{kelkar2} who have reviewed the theoretical basis for the methods used in
the present numerical study. We expect these result will motivate further investigations
parallel to those reported in Ref.~\cite{tshlee}.
  
\begin{figure}[th]
\includegraphics[height=0.55\textwidth, angle=90]{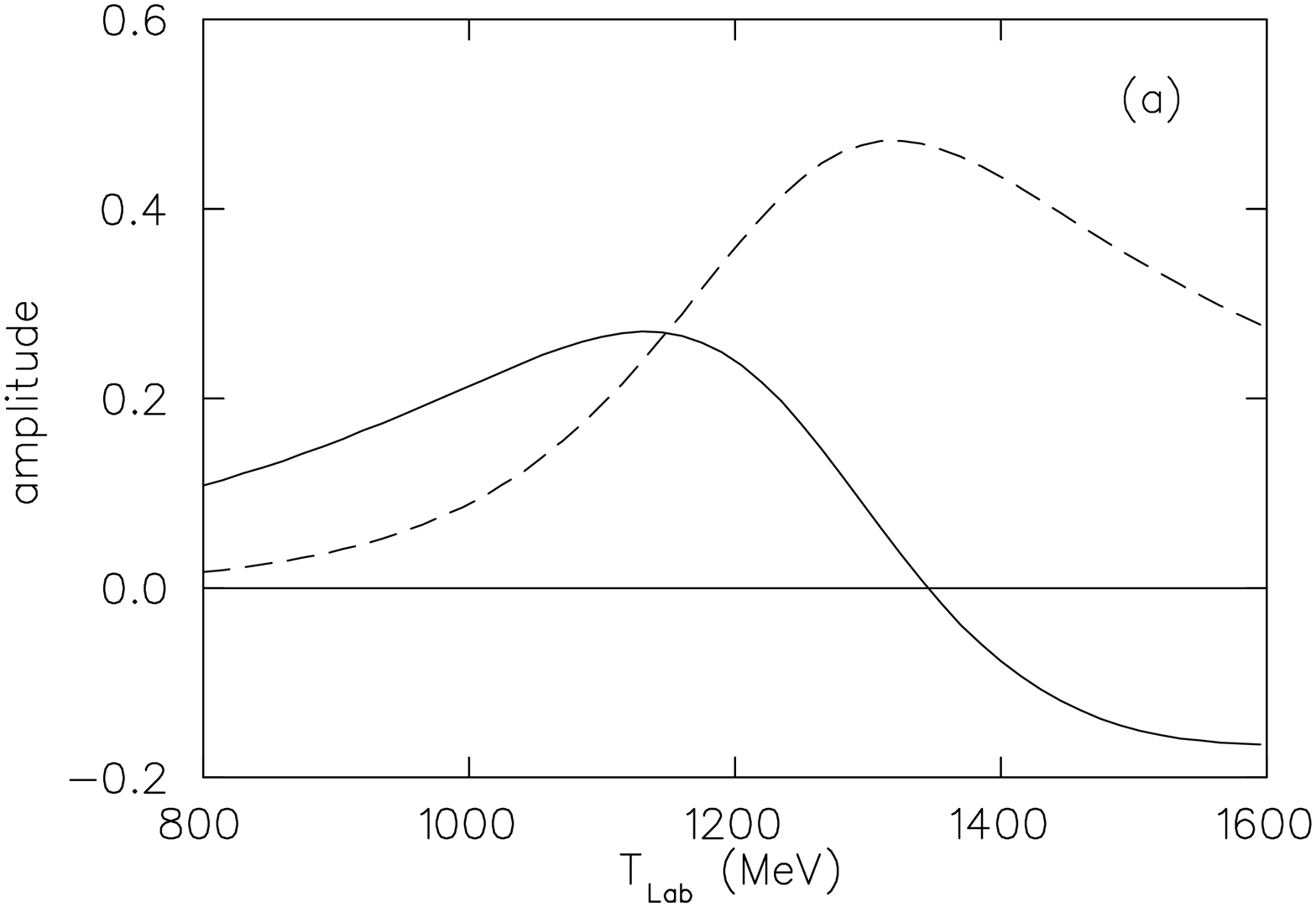}\\
\includegraphics[height=0.55\textwidth, angle=90]{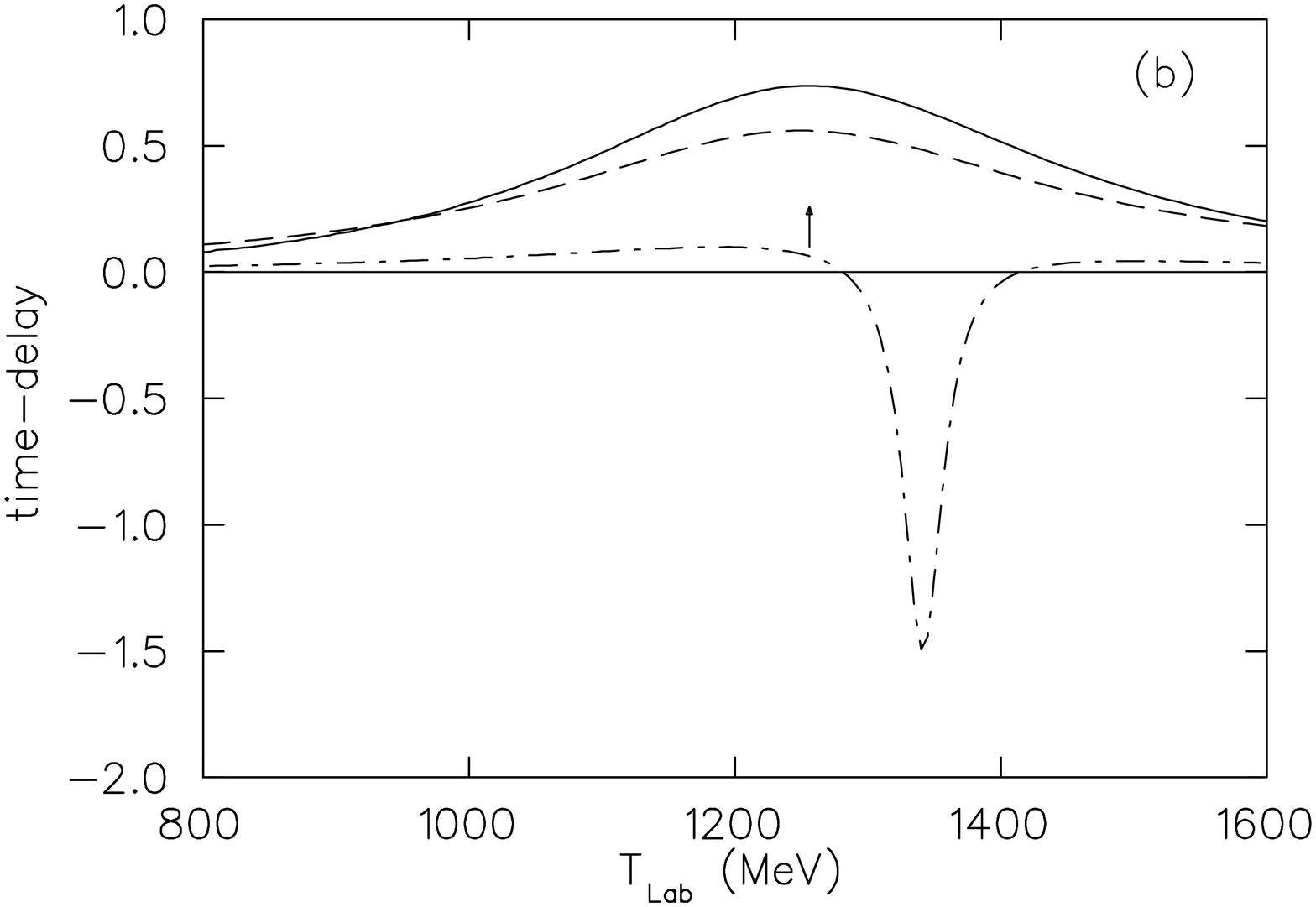}\\
\includegraphics[height=0.55\textwidth, angle=90]{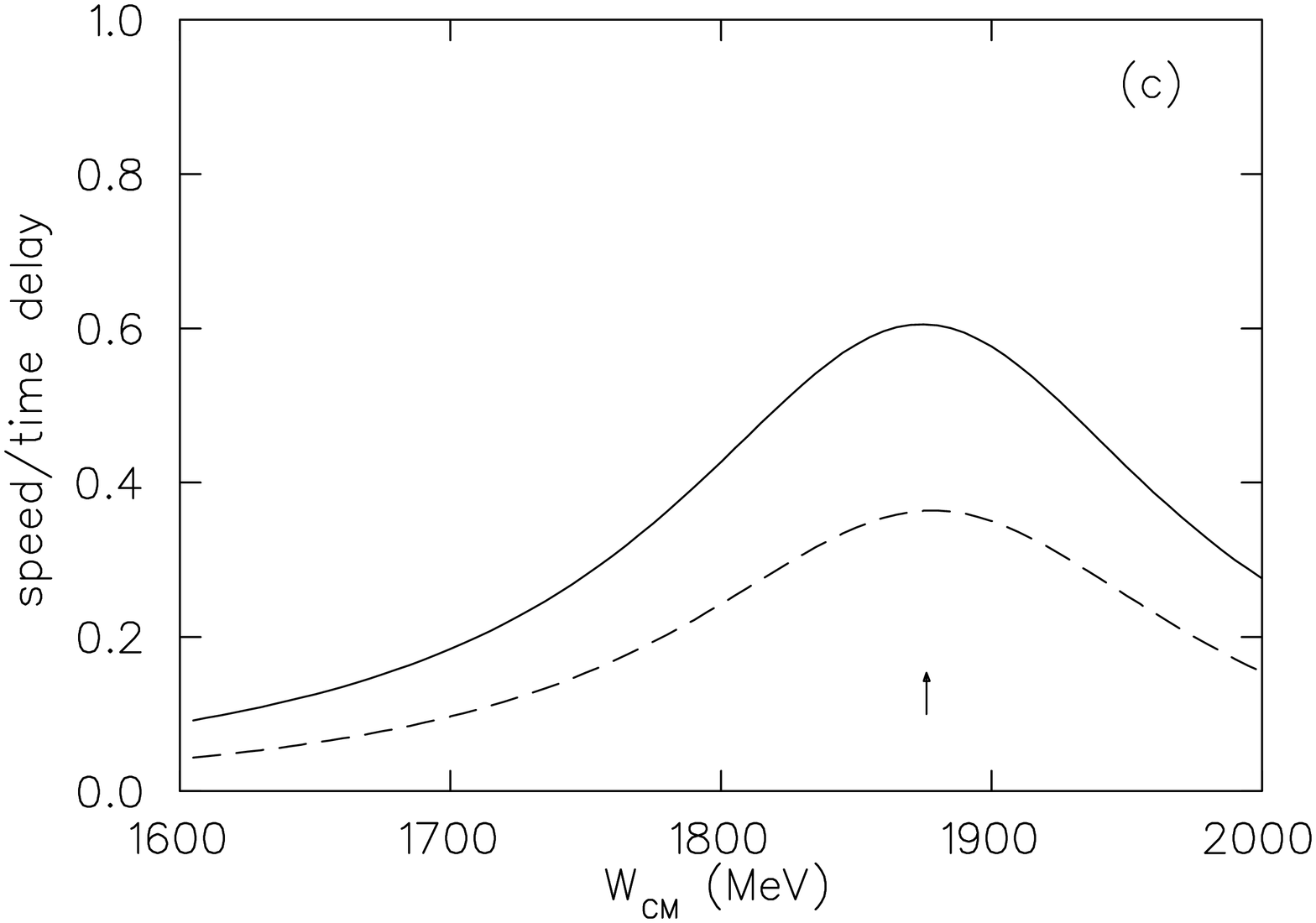}
\caption{Time-delay and speed plots for the F$_{37}$ $\pi N$
resonance $\Delta (1950)$ in arbitrary units. 
Arrows denote the real part of resonance pole
position. Plotted are (a) the real(solid) and imaginary(dashed)
parts of the dimensionless F$_{37}$ $\pi N$ amplitude,
(b) the matrix average time delay of
Smith (dashed) and Ohmura (solid), the result using Eq.~4 (dot-dashed), 
(c) the Eisenbud single-channel
time delay (dashed) and the speed (solid).
}
\end{figure}

\begin{figure}[th]
\includegraphics[height=0.55\textwidth, angle=90]{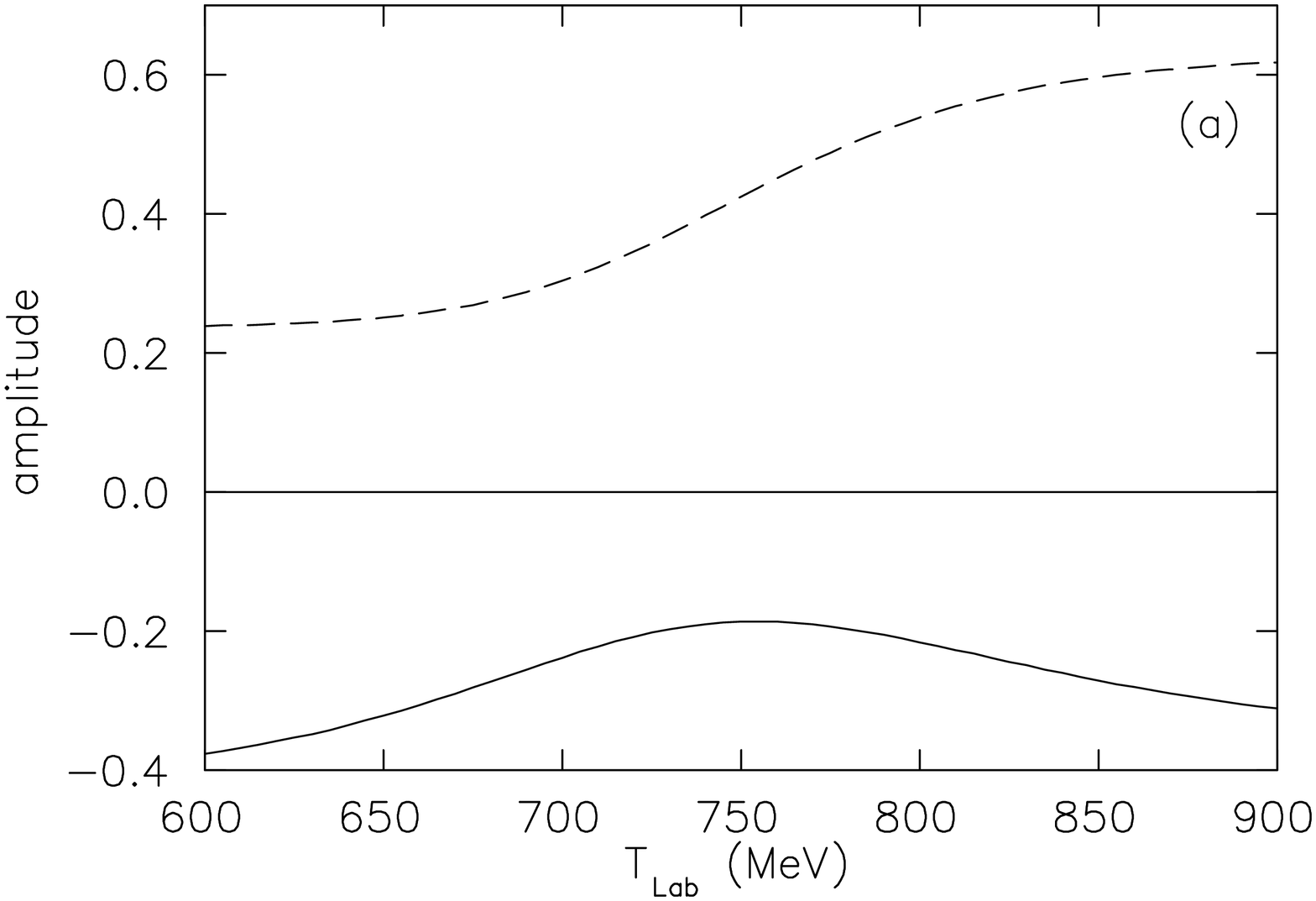}\\
\includegraphics[height=0.55\textwidth, angle=90]{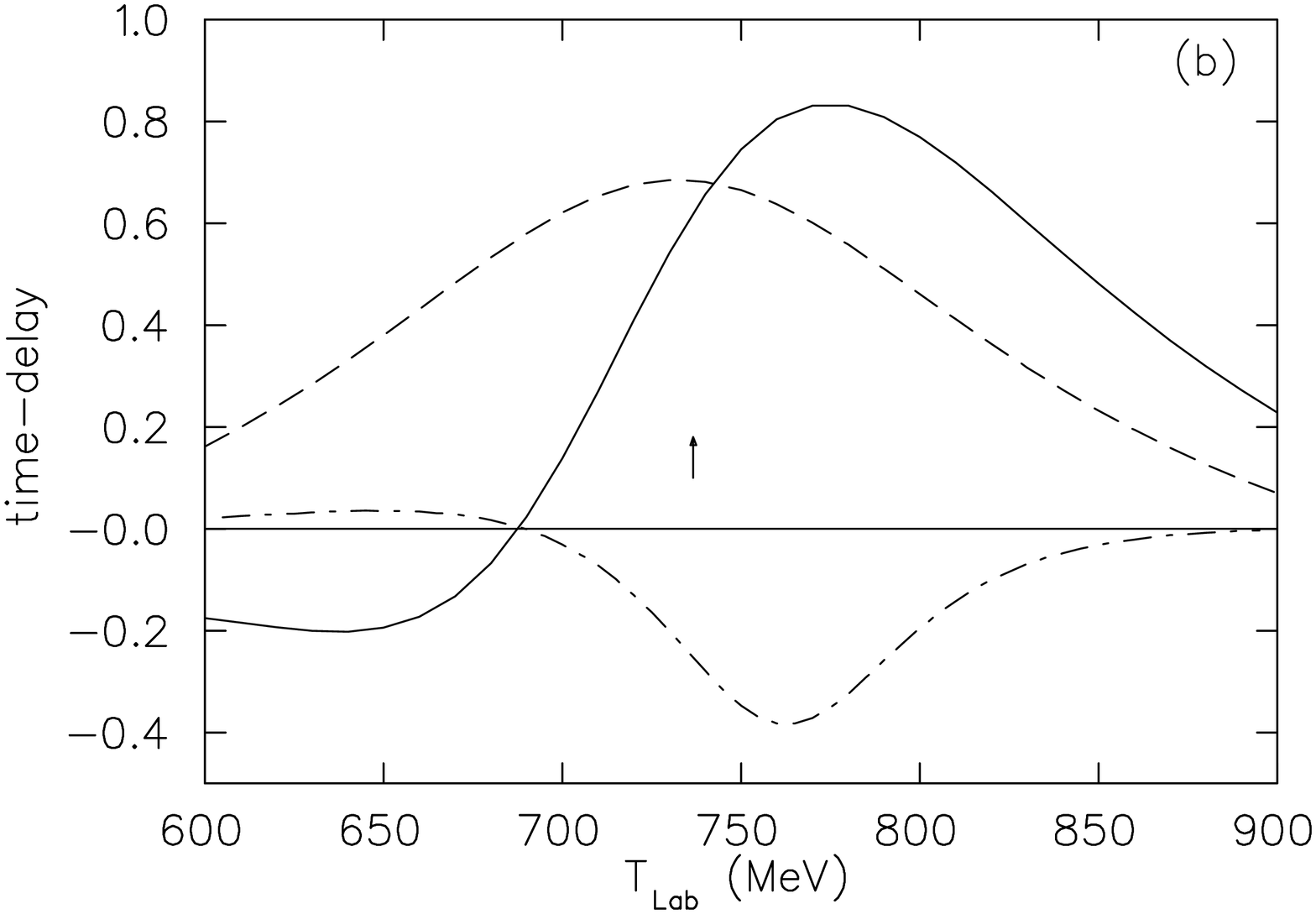}\\
\includegraphics[height=0.55\textwidth, angle=90]{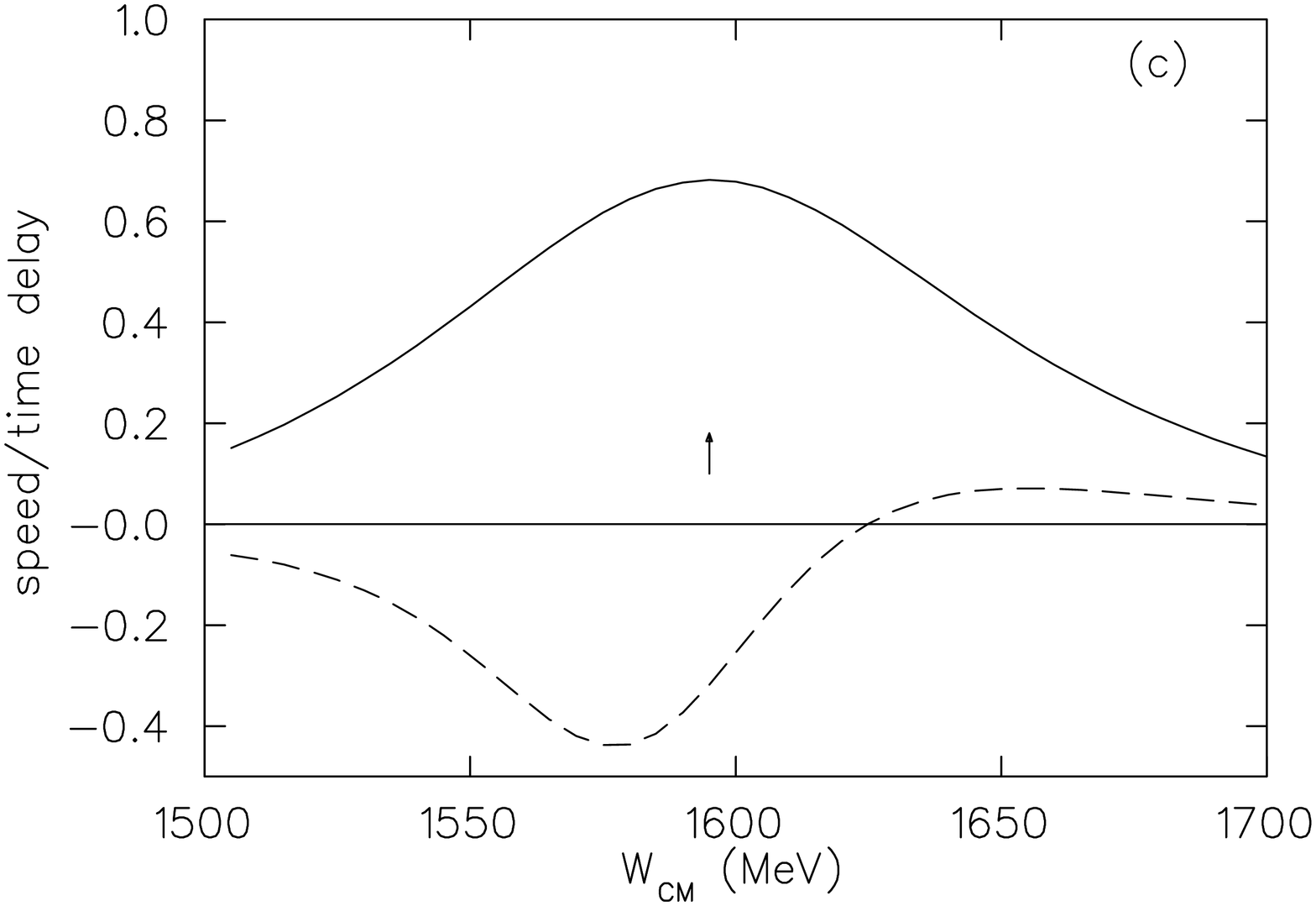}
\caption{Time-delay and speed plots for the S$_{31}$ $\pi N$
resonance $\Delta (1620)$ in arbitrary units.
Arrows denote the real part of resonance pole
position. Plotted are (a) the real(solid) and imaginary(dashed) 
parts of the dimensionless S$_{31}$ $\pi N$ amplitude,
(b) the matrix average time delay of
Smith (dashed) and Ohmura (solid), the result using Eq.~4 (dot-dashed),
(c) the Eisenbud single-channel
time delay (dashed) and the speed (solid).
}
\end{figure}

\begin{acknowledgments}

We thank N.~Kelkar for pointing out the relevance of her recent
work on this topic. 
This work was supported in part by the U.S. Department of
Energy Grant DE-FG02-99ER41110 and funding provided by Jefferson Lab. 
\end{acknowledgments}
\eject

\eject


\begin{thebibliography}{99}
\bibitem{pole}
Pole positions were determined by extrapolating the analytic
expression for each partial-wave amplitude into the complex
energy plane.
\end{thebibliography}
\end{document}